\documentclass[preprint,aps,floats,subeqn,showpacs,amssymb]{revtex4}
\usepackage{epsfig}

\newcommand{\be}{\begin{equation}}
\newcommand{\ee}{\end{equation}}
\newcommand{\bea}{\begin{eqnarray}}
\newcommand{\eea}{\end{eqnarray}}
\newcommand{\bml}{\begin{mathletters}}
\newcommand{\eml}{\end{mathletters}}

\begin{document}

\tighten

\draft




\title{Non-abelian black strings and cosmological constant}
\renewcommand{\thefootnote}{\fnsymbol{footnote}}
\author{F. Bakkali Taheri and Y. Brihaye\footnote{yves.brihaye@umh.ac.be}  }
\affiliation{Faculte des Sciences, Universite de Mons-Hainaut, 7000 Mons, Belgium}

\date{\today}
\setlength{\footnotesep}{0.5\footnotesep}

\begin{abstract}
We study the classical solutions of the Einstein-Yang-Mills
model in five dimensions in the presence of a cosmological constant
$\Lambda$.
Using a spherically symmetric ansatz and assuming that the fields
do not depend on the extra dimension, we transform the equations into
a set of differential equations that we solve numerically.
We construct new types of regular (resp. black holes) solutions
which, close to the origin (resp. the event horizon) resemble
the 4-dimensional gravitating monopole (resp. non abelian black hole)
and study their global properties.
\end{abstract}

 \maketitle
\section{Introduction}
It is believed that topological defects have occured and played a role
during some phase tansitions in the evolution of the Universe, see e.g.
\cite{vilenkin}.
In particular, magnetic monopoles 
\cite{thooft} must have been produced during the GUT
symmetry breaking phase transition. The actual non-observation of them
leads to constaints which have to be implemented in the models of inflation.
On the other hand observational evidences performed in the last years 
\cite{astrocc} favours to possibility that space-time has an accelerated
expansion which could be related to a positive cosmological constant.

It is therefore natural to examine the properties of the various
topological defects in presence of a cosmological constant, or said
in other words, in asymptotically DeSitter space-time.
Recently \cite{bhrds} the magnetic monopole and the sphalerons
occuring in an SU(2)
gauge theory spontaneously broken by a scalar potential were constructed
in an asymptotically DeSitter space-time and it was found that the 
asymptotic decay of the matter field is not compatible with a finite mass.

On the other hand, the last years have seen an increase of
attention for space-times involving more than four dimensions
and of brane-world models to describe space, time and matter.
The so-called brane-world models \cite{brane} have gained a lot of
interest. These assume the standard model fields to be
confined on a 3-brane embedded in a higher dimensional manifold.

A large number of higher dimensional black holes has been studied in
recent years. The first solutions that have been constructed
are the hyperspherical generalisations of well-known black holes
solutions such as the Schwarzschild and Reissner-Nordstr\"om solutions
in more than four dimensions \cite{tan} as well
as the higher dimensional Kerr solutions \cite{mp}.
In $d$ dimensions, these solutions have horizon topology $S^{d-2}$.

However, in contrast to 4 dimensions black holes with different horizon
topologies should be possible in higher dimensions. 
An example is a 4-dimensional
Schwarzschild black hole extended into one extra dimension, a so-called
Schwarzschild black string. These solutions have been discussed extensively
especially with view to their stability \cite{gl}.
A second example, which is important due to its implications for
uniqueness conjectures for black holes in higher dimensions, is the black ring
solution in 5 dimensions with horizon topology $S^2\times S^1$ \cite{er}.
 
The by far largest number of higher dimensional
black hole solutions constructed so far are solutions of the
vacuum Einstein equations, respectively Einstein-Maxwell equations.
The first example of black hole solutions containing non-abelian
gauge fields have been discussed in \cite{bcht}. These are non-abelian
black holes solutions of a generalised 5-dimensional Einstein-Yang-Mills
system with horizon topology $S^3$. 
Using ideas of \cite{volkov,bh1},
SU(2)-black strings with $S_2\times S_1$
topology were constructed in \cite{hartmann}.
Several regular and black hole solutions of an Einstein-Yang-Mills
model have been constructed recently with different symmetries
\cite{bhr,bh2,bch}.
These solutions are non-abelian black hole solutions in 4 dimensions 
extended into one extra codimension.

In this paper, we consider the Einstein-Yang-Mills lagrangian
in five dimensions and gauge group SU(2).
Along with \cite{volkov} we make the hypothesis
that the fields do not depend of the extra dimension
and assume the fields to be  spherically symmetric
in standard three-dimensional space. As shown
in \cite{volkov,bh1} the lagrangian can be dimensionally reduced to
a four-dimensional effective lagrangian describing Yang-Mills-Higgs-dilaton
(the Higgs field being a triplet of SU(2))
interacting minimally with the metric fields.
As so the classical equations admit solutions similar
to the gravitating monopole and monopole black holes
\cite{bfm}, the dilaton plays however a major role and
the pattern of classical solutions is considerably different.

If we extend the original model by means of a 5-dimensional
cosmological constant,
the reduced  effective
action is supplemented by a Liouville potential in the dilaton field
\cite{bbh}.
Several explicit solutions of these equations have been constructed
in the last years \cite{mann,bbh} but, as far as we know, these
solutions present singularities at the origin
 and/or are characterized by non-conventional
asymptotic forms of the metric fields.   Very recently,
a detailed analysis of black string with negative cosmological
constant was reported in \cite{mrs}.
Here we reconsider the classical equations in presence of a positive
cosmological constant, insisting that, close to the origin
(resp. to the event horizon), the fields behave like a gravitating monopole
(resp. a black string).
Such solutions are indeed constructed numerically with the appropriated
boundary conditions and their properties
are analyzed for different values of the coupling constants.

We give the model including the ansatz, equations of motion and
boundary conditions in Section II. The numerical results 
corresponding to solution regular at the origin and solutions
presenting an event horizon
are discussed respectively
in Sections III and IV. The summary is given in Section V.
\section{The Model}

The Einstein-Yang-Mills Lagrangian
in $d=(4+1)$ dimensions is
given by:

\begin{equation}
\label{action}
  S = \int \Biggl(
    \frac{1}{16 \pi G_{5}} (R - 2 \Lambda_5 ) 
    - \frac{1}{4 e^2}F^a_{M N}F^{a M N}
  \Biggr) \sqrt{g^{(5)}} d^{5} x
\end{equation}
with the SU(2) Yang-Mills field strengths
$F^a_{M N} = \partial_M A^a_N -
 \partial_N A^a_M + \epsilon_{a b c}  A^b_M A^c_N$
, the gauge index
 $a=1,2,3$  and the space-time index
 $M=0,...,5$. $G_{5}$, $\Lambda_5$ and $e$ denote
respectively the $5$-dimensional Newton's  and cosmological
constants and the coupling
constant of the gauge field theory. $G_{5}$ is related to the Planck mass
$M_{pl}$ by $G_{5}=M_{pl}^{-3}$ and $e^2$ has the dimension of 
$[{\rm length}]$.

Along with \cite{volkov} we assume  the metric and the matter fields
to be  independent on the extra coordinate $y$, the fields can be
parametrized as follows :
\begin{equation}
g^{(5)}_{MN}dx^M dx^N = 
e^{-\xi}g^{(4)}_{\mu\nu}dx^{\mu}dx^{\nu}
+e^{2\xi} dy^2 
\ , \ \mu , \nu=0, 1, 2, 3 
\end{equation}
and
\begin{equation}
\label{gauge}
A_M^{a}dx^M=A_{\mu}^a dx^{\mu}+ \Phi^a dy   \  . \
\end{equation}
$g^{(4)}$ is the $4$-dimensional metric tensor.
After  dimensional reduction,
the lagrangian above leads
to an effective 4-dimensional Einstein-Yang-Mills-Higgs-dilaton
lagrangian whose matter part $L_M$ is reads~:
\begin{equation}
L_M = -\frac{1}{4} e^{ \xi} F^a_{\mu \nu}F^{a,\mu \nu}
      -\frac{1}{2} e^{-2 \xi} D_{\mu} \Phi^a D^{\mu} \Phi^a
      - \frac{3}{4} \frac{1}{\alpha^2} \partial_{\mu} \xi \partial^{\mu} \xi
      - \frac{\Lambda}{2} e^{- \xi} \ \ , \ \ \alpha^2 \equiv 4 \pi G
\end{equation}
The cosmological constant in $d=5$ leads to a Liouville
potential for the dilaton \cite{bbh} in $d=4$.

\subsection{The Ansatz}
Our aim is to construct non-abelian regular and black strings
solutions  which are
spherically symmetric in the four-dimensional space-time
and are extended
into one extra dimension. The topology of these non-abelian
black strings will thus be $S^2\times \mathbb{R}$ or
 $S^2\times S^1$ if   the extra coordinate
$y$ is chosen to be periodic.
 
For the metric the spherically symmetric Ansatz reads:
\begin{equation}
g^{(5)}_{MN}dx^M dx^N =
e^{-\xi}\left[-A^{2}Ndt^2+N^{-1}dr^2+r^2 d\theta^2+r^2\sin^2\theta
d^2\varphi\right]
+e^{2\xi} dy^2 
\label{metric}
\ , \end{equation}
where $N,A, \xi$ are function of the coordinate $r$ only. As usual we define
\begin{equation}
N(r)=1-\frac{2m(r)}{r} - \frac{\Lambda}{6} r^2 \ \ {\rm or} \ \
  N(r)=1-\frac{2 \tilde m(r)}{r}
\ . \end{equation}
In these coordinates, $m(r)$ represents the
(dimensionful) mass per unit length (of the extra dimension)
of the solution.

For the gauge fields, we use the spherically symmetric ansatz
\cite{thooft} :
\begin{equation}
\label{ansatz1}
{A_r}^a={A_t}^a=0
\ , \end{equation}
\begin{equation}
\label{ansatz2}
{A_{\theta}}^a= (1-K(r)) {e_{\varphi}}^a
\ , \ \ \ \
{A_{\varphi}}^a=- (1-K(r))\sin\theta {e_{\theta}}^a
\ , \end{equation}
\begin{equation}
\label{higgsansatz}
{\Phi}^a=v H(r) {e_r}^a \ \ , 
\end{equation}
where $v$ is a mass scale.

\subsection{Equations of motion}
With the following rescalings:
\begin{equation}
x=evr \ \ , \ \ \mu=evm
\end{equation}
the resulting set of ordinary differential equations only depends
on the fundamental coupling $\alpha \equiv 4\pi \sqrt{G_{5}} v$
and of the reduced cosmological constant $\Lambda \equiv 2 \alpha^2 \Lambda_5$.

The Einstein equations for the metric functions $N$, $A$ and $\xi$ then read:

\begin{eqnarray}
\label{eqgra}
\mu ' &=& \alpha^2 \left(e^{\xi}N(K')^2 + \frac{1}{2}N x^2(H')^2
e^{-2\xi}+
\frac{1}{2x^2}(K^2-1)^{2} e^{\xi}+K^2 H^2 e^{-2\xi}\right) \nonumber 
\\
&+& \frac{3}{8}Nx^{2}(\xi ')^2  + \frac{\Lambda}{4} x^2 e^{- \xi} \ ,
\end{eqnarray}
\begin{equation}
A'=\alpha^2 x A \left(\frac{2(K')^2}{x^2}e^{\xi}+
e^{-2\xi}(H')^2\right)+\frac{3}{4} x A(\xi ')^2
\ , \label{dgl5} \end{equation}
\begin{eqnarray}
(x^2 AN\xi')' &=& \frac{4}{3}\alpha^2
A \left[e^{\xi}\left(N(K')^2+\frac{(K^2-1)^2}{2
x^2}\right)- 2
e^{-2\xi}\left(\frac{1}{2}
N (H')^2 x^2+H^2 K^2\right) \right]  \nonumber
\\
&-&  \frac{\Lambda}{3} A x^2 e^{- \xi}
\ , \label{dgl3} \end{eqnarray}
while the Euler-Lagrange equations for the matter 
functions $K$ and $H$ are given by:
\begin{equation}
(e^{\xi}ANK')'=A\left(e^{\xi}\frac{K(K^2-1)}{x^2}+e^{-2\xi}H^2 
K\right)
\ , \label{dgl1} \end{equation}
\begin{equation}
(e^{-2\xi}x^2 ANH')'=2e^{-2\xi}K^2 AH
\ , \label{dgl2} 
\end{equation}
where the prime denotes the derivative with respect to $x$.

These equations are invariant under the dilatation transform 
\begin{equation}
  e^{\xi} \to \lambda e^{\xi} \ , \ 
  x \to \lambda^{3/2} x \ , \
  \alpha^2 \to \lambda^2 \alpha^2 \ , \
  \Lambda \to \lambda^{-2} \Lambda \ , \
  \mu \to \lambda^{3/2} \mu
\end{equation}
where $\lambda$ is an arbitrary constant. In \cite{bh1} the "gauge"
$\xi(\infty)=0$ was chosen. Here, the symmetry above will be
fixed by imposing  $\xi(0)=0$.  The pattern
of solutions is  quite different with this choice, 
in particular solutions
exist for arbitrary values of $\alpha$ as indicated on Fig. 1 which has
to be contrasted with Figs.1 and 3 of \cite{bh1}.
In order to avoid confusion with previous papers on the topic,
we will use $\alpha, x_h$ to denote the gravitation coupling
constant and the black hole horizon in the gauge $\xi(\infty)=0$
and $\alpha', x_h'$  to denote these quantities in the gauge
$\xi(0)=0$.

\subsection{Boundary conditions}
The boundary conditions for a regular solution at the origin read
\begin{equation}
\label{bc0}
 \mu(0)=0 \ \ , \ \ K(0) = 1 \ \ , \ \ H(0)=0 \ \ , \ \ \xi'(0)=0
\end{equation}
which have to supplemented with our gauge fixing $\xi(0)=0$.
At infinity, finiteness of the mass and asymptotic
flatness of the four-dimensional space-time requires:
\begin{equation}
\label{bcinfty}
A(\infty)=1 \ \ , \ \ K(\infty)=0 \ \ , \ \ H(\infty)=1 \ \ ,
\ \ \xi '(\infty)=0 \ .
\end{equation}
Note that the function $A(r)$ can in fact be rescaled at will and can
be normalized according to $A(r_0)=1$ at any point $r_0 \in [0,\infty]$

The boundary conditions at the regular horizon $x=x_h$ read:
\begin{equation}
\label{bc1}
N(x_h)=0  \Rightarrow \tilde  \mu(x_h)=\frac{x_h}{2}
\end{equation}
with $A(x_h) < \infty$ and
\begin{equation}
\label{bc2}
(N' K')|_{x=x_h}=\left[\frac{K(K^2-1)}{x^2}+e^{-3\xi} H^2 K\right]\vert_{x=x_h} \ ,
\end{equation}
\begin{equation}
\label{bc3}
(N' H')|_{x=x_h}=\left(\frac{2}{x^2}K^2 H\right)\vert_{x=x_h} \ ,
\end{equation}
\begin{equation}
\label{bc4}
(N' \xi')|_{x=x_h}=\left\{\frac{4}{3x^2}\alpha^2\left[e^{\xi}\left(N(K')^2+\frac{(K^2-1)^2}{2
x^2}\right)- 2
e^{-2\xi}\left(\frac{1}{2}
N (H')^2 x^2+H^2 K^2\right) \right] \right\} \vert_{x=x_h} \ \ .
\end{equation}
 In this case as well we will choose the dilatation symmetry by demanding
 $\xi(x_h)=0$.

One of the main results of our numerical analysis
is that, in the case $\Lambda > 0$, the solution ends up
at a cosmological horizon $x=x_c$ where $N(x_c)=0$. The
behaviour of the other metric fields for $x \to x_c$ suggest that
the system of coorddinates used in not appropriate
 for extending the  solutions for $x > x_c$.
As a consequence, it is impossible to impose a set of conditions of the
type (\ref{bc4}) at $x=x_c$.
Appropriate boundary conditions have therefore
to be enforced in the case $\Lambda > 0$. We will consider only
the "realistic case" $\Lambda << 1$. In this case, at least
the matter functions reaches
their asymptotic values $K=0,H=1$ much before the cosmological horizon
and can be imposed in the limit $x \to x_c$.
The six other boundary conditions have to be imposed at the origin
(or at the event horizon in the case of black strings).

\section{Regular solution at the origin}
\subsection{ Case $\Lambda = 0$}
We solved the equations (\ref{eqgra})-(\ref{dgl3}) subject to the conditions
(\ref{bc0}) and (\ref{bcinfty}) (or  (\ref{bc4}) in the case of
black holes) by numerical method
for several values of  $\alpha'$.
It turns out that, in the gauge chosen,
 solutions exist for a large domain of $\alpha'$ which we believe
 is infinite. Some characteristics are presented on Fig. 1;
 in particular the asymptotic value $\xi(\infty)$ and the charge
 of the dilaton say $Q_d$ defined according to $\xi \approx \xi(\infty) + Q/x$.

\subsection{ Case $\Lambda > 0$}
When the cosmological constant parameter is choosen to be positive,
 the function $N(x)$ develops (as expected) a zero
at some finite value of the coordinate $x$, say $x=x_c$,
this is called a cosmological horizon.
 Indeed, for $\vert x_c - x \vert << 1$, we find the behaviour
\begin{equation}
     N(x) \sim  N_c(x_c-x) \ \ , \ \
\xi(x) \sim \xi_i +  \xi_c \sqrt{ (x_c-x)}   \ \ , \ \
    A(x) = A_c (x_c-x)^{-a}
\end{equation}
where $N_c, \xi_i,  \xi_c, A_c, a$ are constants (with $a > 0$)
  depending on $\alpha$ and $\Lambda$. 
The best we can do is to integrate the equations on $x \in [0, x_c]$.
For this purpose, we considered only values of $\Lambda$ such that
the gauge and Higgs fields attain their asymptotic values for $x << x_c$.
This is possible because $x_c \to \infty$ when $\Lambda \to 0$.
So for sufficiently small values of the cosmological constant one is
 to impose $K(x_c)=0, H(x_c)=1$ as boundary conditions;
all other conditions being  supplemented at $x=0$
 (see (\ref{bc0})).
 The solution corresponding to $\alpha=1,\Lambda = 0.0001$ is presented
 on Fig. 2; it corresponds to $x_c \approx 183$. The profiles of the
 corresponding solution for vanishing cosmological constant are also
 reported. We see how the cosmologic solution deviates from the asymptotically
 flat one at sufficiently large values of $x$.
 On the figure we can see that the function $a(r)$ develops a plateau
 at intermediate values between $r=0$ and $r=r_c$. We found it convenient
 to normalize this function in such a way that it reaches the
 unit value on the plateau.

 It is natural to study the evolution of the different parameters 
 characterizing  the solution at $r=r_c$ for different values of
 $\alpha'$ and of $\Lambda$. The analysis is not so easy because the
 singular point has to be approached closer and closer. However we
 performed such an analysis for the case  $\Lambda = 0.0001$. The
 values of $x_c$, $\xi(x_c)$, $N'(x_c)$ and $a$ are reported of Fig.3.
 In particular, we observe that for $\alpha' > 1$ both values 
 $x_c$ and $\xi(x_c)$ increase for increasing $\alpha'$. This can be
 related to the fact that the combination $e^{-\xi(x_c)} \Lambda$
 determines an effective
 coupling constant which for instance become smaller for increasing $\alpha'$.
 Explaining the fact that the value of the cosmological horizon is
 pushed towards infinity.
 Remember that, having fixed $\xi(0)$ as a boundary condition, we have
 no control on $\xi(x_c)$. 
 The figure further shows that the values $a$ and $N'(x_c)$
 vary only a little with $\alpha'$. The figure is limited to $\alpha' \leq 1.5$;
 the numerical analysis becomes more difficult with increasing $\alpha'$,
 likely because both, the minimum $N_m$ and the value $A(0)$ get smaller
 and smaller.
 The profiles of the metric functions available for $\alpha'=1.5$ are presented
 on Fig.4 and illustrate our claim. 
 By comparing Figs. 2 and 4, we can also appreciate
 the evolution of the dilaton function. It turns out that this function
 develops a local minimum in the neighborood of the radius where $N$ gots
 its minimum.
 We believe that Figs. 2-4 reflect the qualitative properties of a generic
  $\Lambda$.

It is a natural question to determine whether the singularity
occuring at $x=x_c$ is essential or just a coordinate singularity.
The computation of the Ricci scalar contains several terms of
 second order in the functions $N$ and $A$, some of them
are apparently singular but the complexity of this expression
does not make easy to conclude. However, taking the trace of the
Einstein equations leads to 
\be
    R = \frac{16 \pi G}{2-d} T - \frac{2 d}{2-d} \Lambda \ \ , \ \ 
    T \equiv T^A_A  \ \ .
\ee
Further noticing  that $T = -(1/2) {\cal L}_{ym}$ for five-dimensional gauge
field theory, it turns out easy to evaluate $R$ in terms of the lagrangian
density ${\cal L}_{ym}$. Since the matter fields reach their expectation
value for $x \to x_c$, it becomes obvious that $R$ is finite for this 
limit. This suggests (although it is not a proof) that $x=x_c$ is a coordinate
singularity and that the problem could be eliminated with  more appropriate
coordinates.
\section{Black strings}
 \subsection{ Case $\Lambda = 0$}
Before presenting our results, let us briefly recall 
how the pattern of solutions
looks like \cite{hartmann,bh2,bhr} in the gauge
$\xi(\infty)=0$. The black string solutions exist
in a limited domain
of the $\alpha$-$x_h$-plane. For fixed value of $x_h$ several branches
of solutions
exist when varying $\alpha$. This is very similar to the globally regular
counterparts \cite{volkov}. When $\alpha$ is fixed and $x_h$ varied, always
two branches of black string solutions exist. The lower branch extends from
the corresponding globally regular solution with lowest 
energy up to a maximal value of
$x_h(\alpha)$. The second branch either terminates 
at a finite $x_h$ and joins
the branch of Einstein-Maxwell-dilaton solutions or extends all the way back
to $x_h=0$, where it joins the branch of globally 
regular solutions with highest energy.

Again, the pattern of black string solutions looks completely different
in the gauge $\xi(0)=0$. For instance, the domain is not limited in the
$\alpha'$ direction. For $\alpha' > \alpha'_c$ (with $\alpha'_c \sim 1.0$)
we observe that   there exist only one  branch of solutions, indexed
by $x'_h$ and which bifurcates into an Einstein-Maxwell-dilaton solution
at some maximal ($\alpha$-depending) value $x'_{h,max}$.
The sketch of the domain is presented in Fig. 5 
; this can be contrasted with Fig. 3 of \cite{bhr}
and on Fig. 6
different parameters characterizing the black string are plotted
as functions of $x'_h$ for $\alpha'=0$.
For $\alpha' < 1.0$ our numerical analysis indicates that the black strings
solutions exist up to a maximal value as well but, then, another branch
of solution exist while decreasing $x'_h$. Only on the second branch
does the black string bifurcate into an abelian solution.
 \subsection{ Case $\Lambda > 0$}
Solutions behaving like black strings for $r << \infty$ and with a
positive cosmological constant can be constructed numerically.
It can be shown that that are plagued with the same singularity 
as their regular-at-the-origin conterparts (see Sect. IV.B).
  The profile of such a solution corresponding to $\alpha = 1, x_h = 0.3$
  is presented on Fig. 7.

\section{$\Lambda > 0$ Black strings - vacuum case}
After observing the recurent pathologies of the non abelian solutions 
in the presence of a positive cosmological constant, we turned
out to the case where matter is absent.  In the case of a negative
cosmological constant, the solutions have been constructed in \cite{mrs}.

If we set $\alpha = 0$ the matter fields equations above
are trivially satisfied by $K=1,H=0$ and we are left with
a system of three Einstein equations for $m(r),A(r), \xi(r)$.
The cosmological constant $\Lambda$ can be absorbed by means
of a translation of $\xi$. Solving these equations with the
regular conditions at the origin (or the ones of an event horizon at $r=r_c$)
leads to the same pathologies as observed in the previous section.
We therefore integrated the equations by imposing the regular conditions
at the cosmological horizon and then integrating from $r=r_c$. Because
of the scale invariance of the system, the value $r_c$ can be 
chosen arbitrarily
and we set $r_c=10$. An extra translation of the function $\xi$ allows one
to fix the boundary condition "a la Cauchy",i.e.
\begin{equation}
                m(r_c) = \frac{r_c}{2} \ \ , \ \ \ A(r_c)=1 \ \ , \ \
                \xi(r_c)=0 \ \ , 
                \ \ \xi '(r_c) = - \frac{2 \Lambda r_c}{3(2-r_c^2)}
\end{equation}
where the last condition ensures the regularity at the horizon.

Integrating the equations  inside
 the horizon, our numerical analysis indicates that the solution
become singular at the approach of the origin. For instance we find~:
\begin{equation}
              m(r) \sim M_0 r^{\omega} \ \ , \ \
              \xi(r) \sim  \xi_0 + c \log(r) \ \ , \ \
              A(r) \sim  A_0 r^{3 c^2/4}
\end{equation}
where $M_0,A_0,c$ and $\omega$ are constants. 
Such a behaviour is confirmed by the asymptotic analysis of the equations.
The power $\omega$ and the other coefficients are determined numerically.
In the case $\Lambda=0.1$
we find
$M_0 \approx -28.057$,
$\omega \approx - 0.7585$,
$\xi_0 \approx -2.1579$,
and $c \approx 1.0056$.
 The profiles of the functions are represented on Fig.1.
 Comparaison with the non Abelian solutions discussed in the
 previous section shows how much the gauge fields regularize the
 solutions at the origin.

Integrating the equations outside the sphere determined by the
cosmological horizon
leads to a behaviour similar to the one above.
The corresponding parameters are $\omega \approx 2.31, c \approx 0.67$.
The asymptotic solution is clearly not DeSitter.
We see that the occurence of a positive cosmological constant has
important effects on the black string solutions and leads to  
completely different
properties than in the case of a negative cosmological constant \cite{mrs}.
Notice that these solutions presented here are different from the
 analytical solutions reported in Sect. 4 of
 \cite{bbh} which have no cosmological
 horizon. A systematic study of vacuum black string solutions
 with positive cosmological constant
 and in arbitrary dimensions will be presented elsewhere \cite{brs}.


\section{Summary}
 We investigated the most natural extention of the gravitating monopole
 living in a five-dimensional space-time endowed with a cosmological constant.
  For technical reasons, we were forced to reconstruct the
 non-abelian black strings in another gauge where the dilaton is imposed
 to be zero at the origin (or at the regular event horizon $x_h$ in the case
 of black strings).
 Although equivalent to the case of the $\xi(\infty)=0$-gauge (chosen e.g. in
  \cite{hartmann,bhr}),
 the pattern of black strings solutions looks completely different.
 We have characterized the domain of solutions in this new frame.

 We further found that supplementing the five dimensional space-time
 by a positive cosmological constant
  leads to a cosmological horizon which constitutes
   an apparent singularity of the solution. 
   The situation, however contrasts to the four dimensional
   case where all functions remain analytic
   when the cosmological horizon is approached
    \cite{bhrds};
   this feature is typically due to the occurence
   of a dilaton field.   Considering the same equations in absence of
   gauge and Higgs fields, our results strongly suggest that the
   presence of a positive cosmological constant does not deform
   smoothly the uniform black strings and leads to solution which
   present a singularity
   at the  origin and whose asymptotic behaviour is not DeSitter.
\\
\\
\\
{\bf\large Acknowledgements} \\
Y. B. thanks  the
Belgian FNRS for financial support
 and gratefully acknowledges T. Delsate,
Eugen Radu and Cristian Stelea for
valuable discussions.

\newpage
\begin{figure}
\centering
\epsfysize=15cm
\mbox{\epsffile{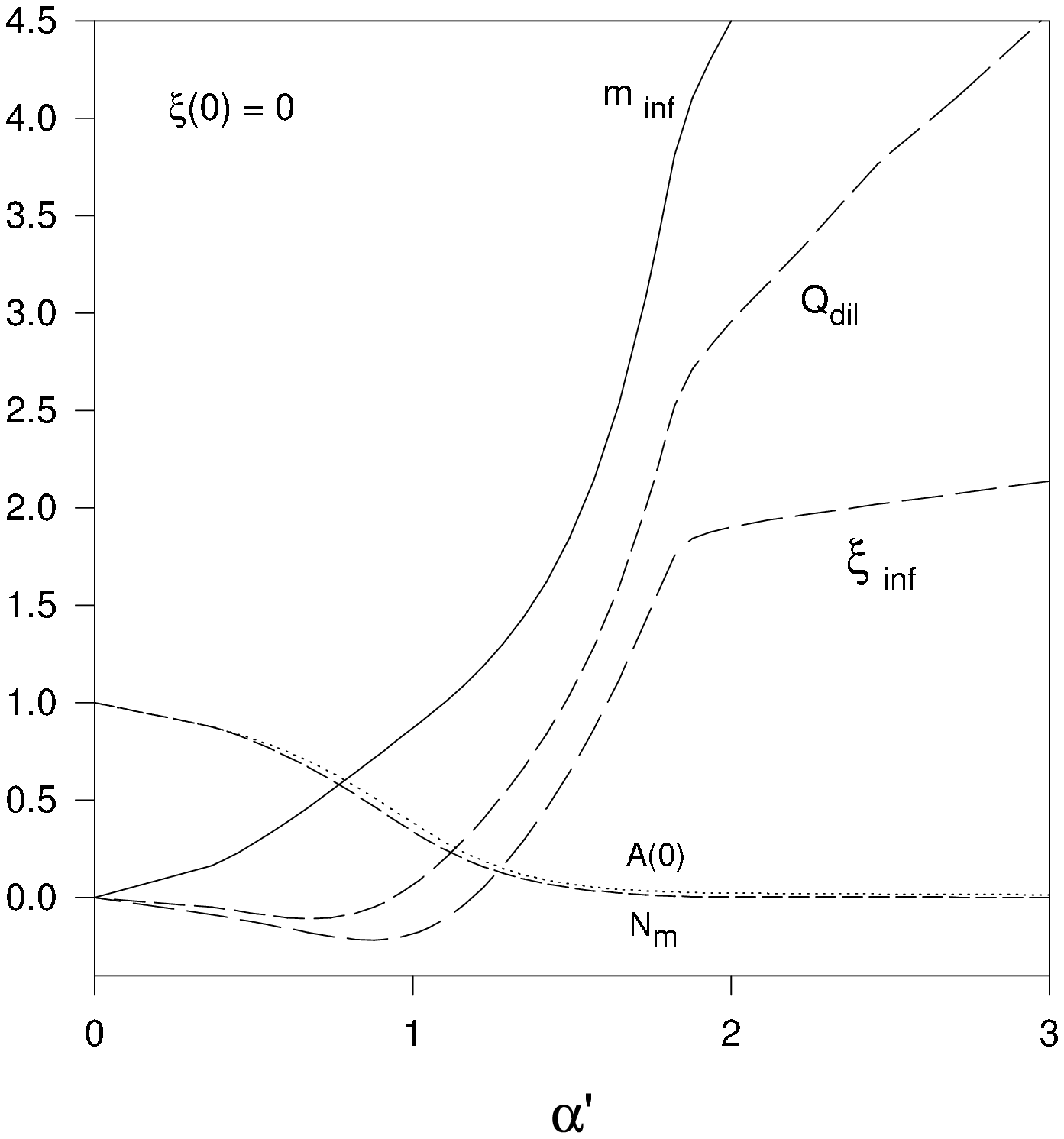}}
\caption{\label{fig1}
The values of the mass, dilaton charge, $A(0), N_{min}$
and $\xi(\infty)$ are plotted as functions of $\alpha_c$
for the solutions with $\xi(0)=0$}
\end{figure}
\begin{figure}
\centering
\epsfysize=17cm
\mbox{\epsffile{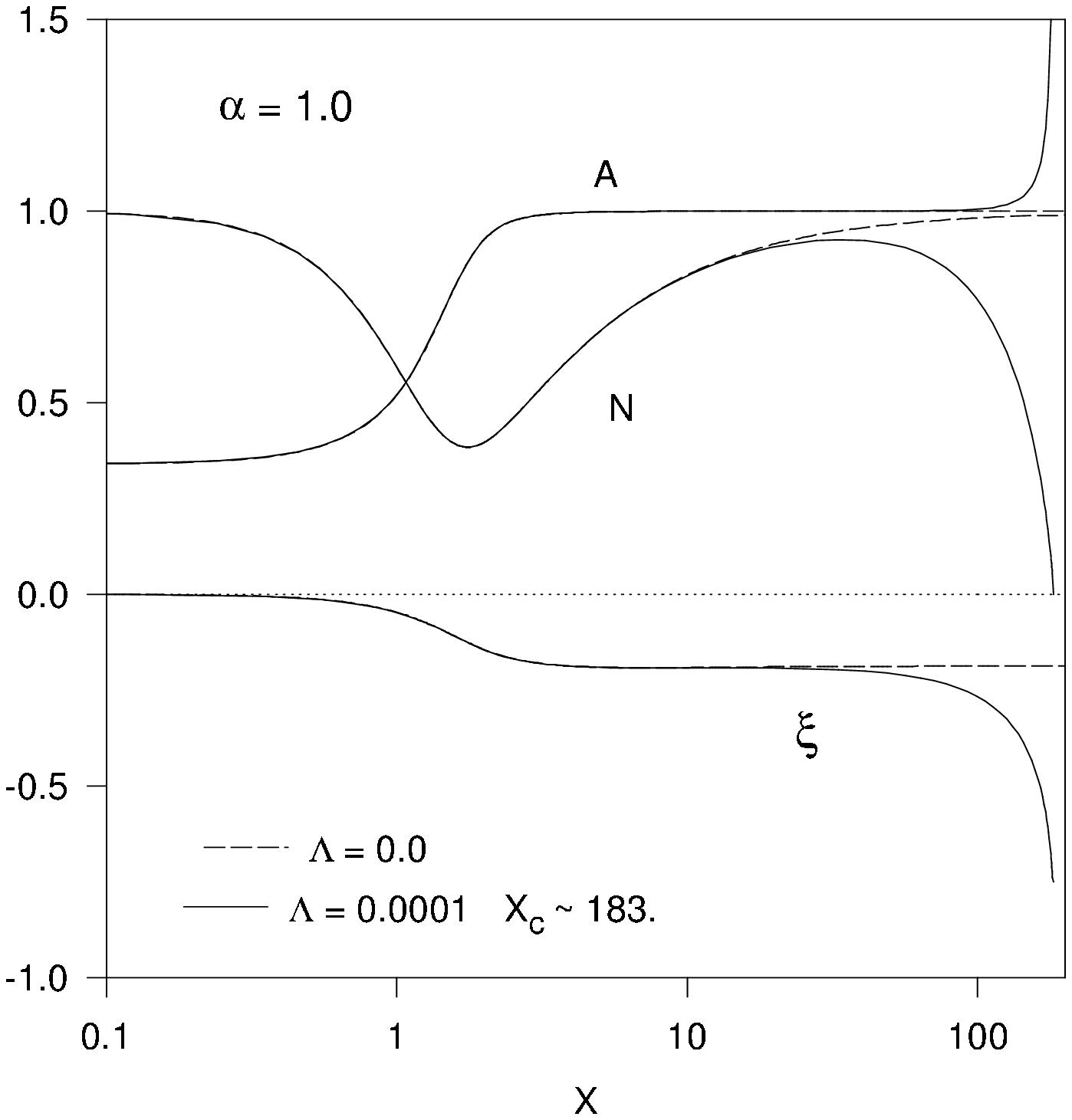}}
\caption{\label{fig2} 
The profile of the metric functions corresponding to
  $\alpha_c=1$ for  $\Lambda = 0.0$ and $\Lambda = 0.0001$. }
\end{figure}
\begin{figure}
\centering
\epsfysize=17cm
\mbox{\epsffile{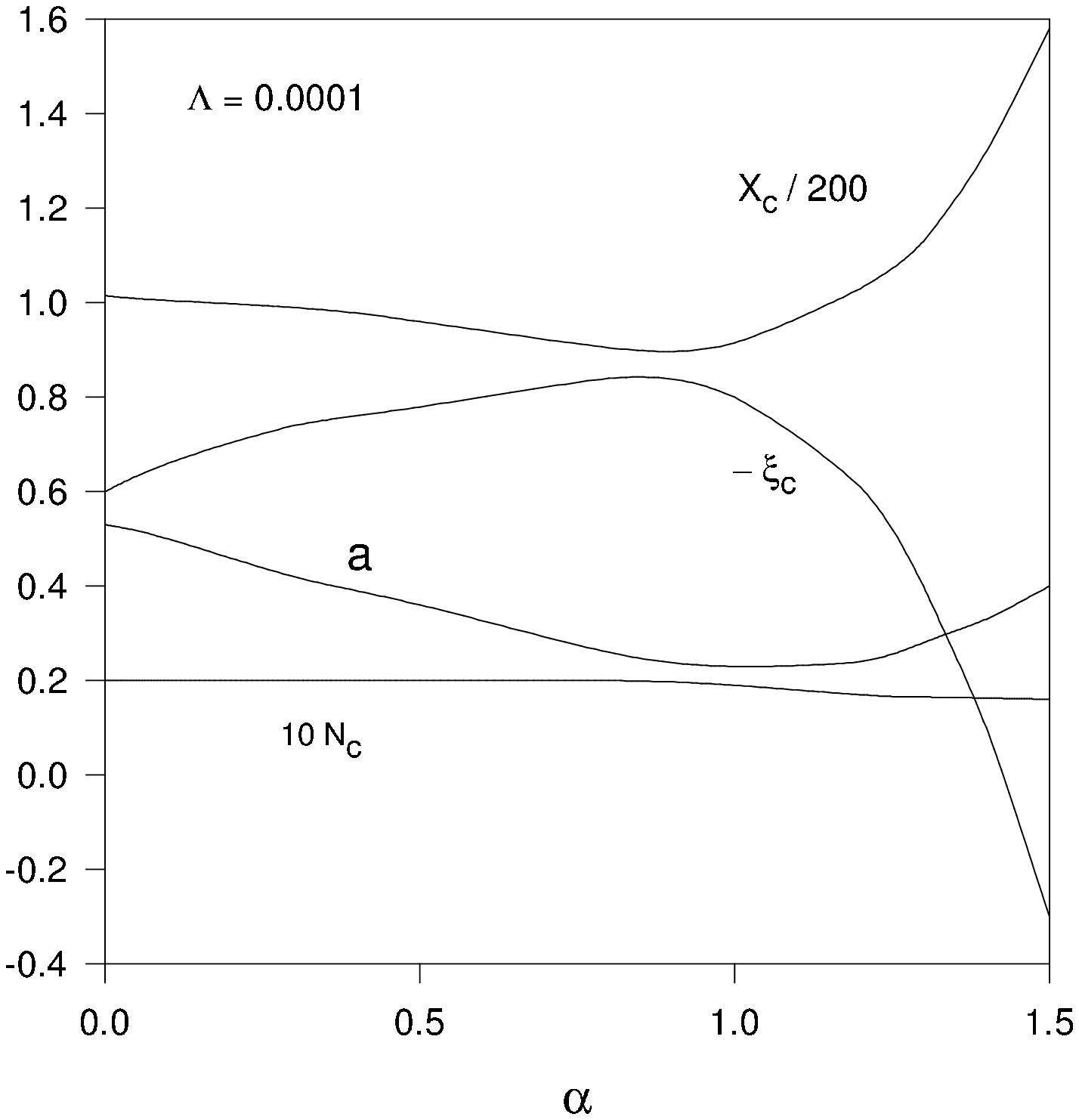}}
\caption{\label{fig3}
The evolution of the parameters at the horizon are shown as functions
of $\alpha$ for $\Lambda = 0.0001$. }
\end{figure}
\begin{figure}
\centering
\epsfysize=17cm
\mbox{\epsffile{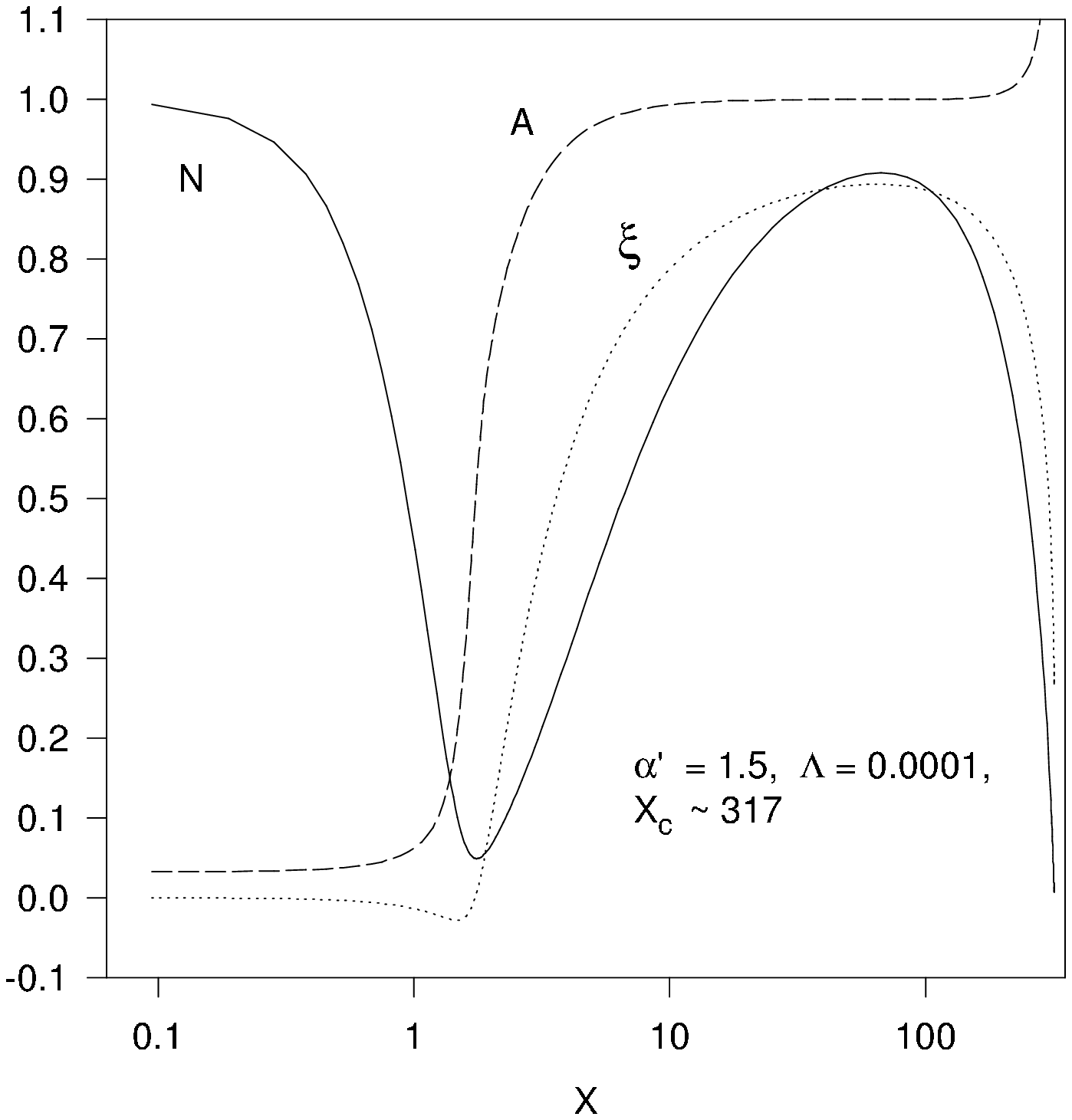}}
\caption{\label{fig4}
Idem fig.2 for $\alpha=1.5$. }
\end{figure}

\newpage
\setlength{\unitlength}{1cm}
\begin{picture}(6,9)
\centering
\put(1.6,0){\epsfig{file=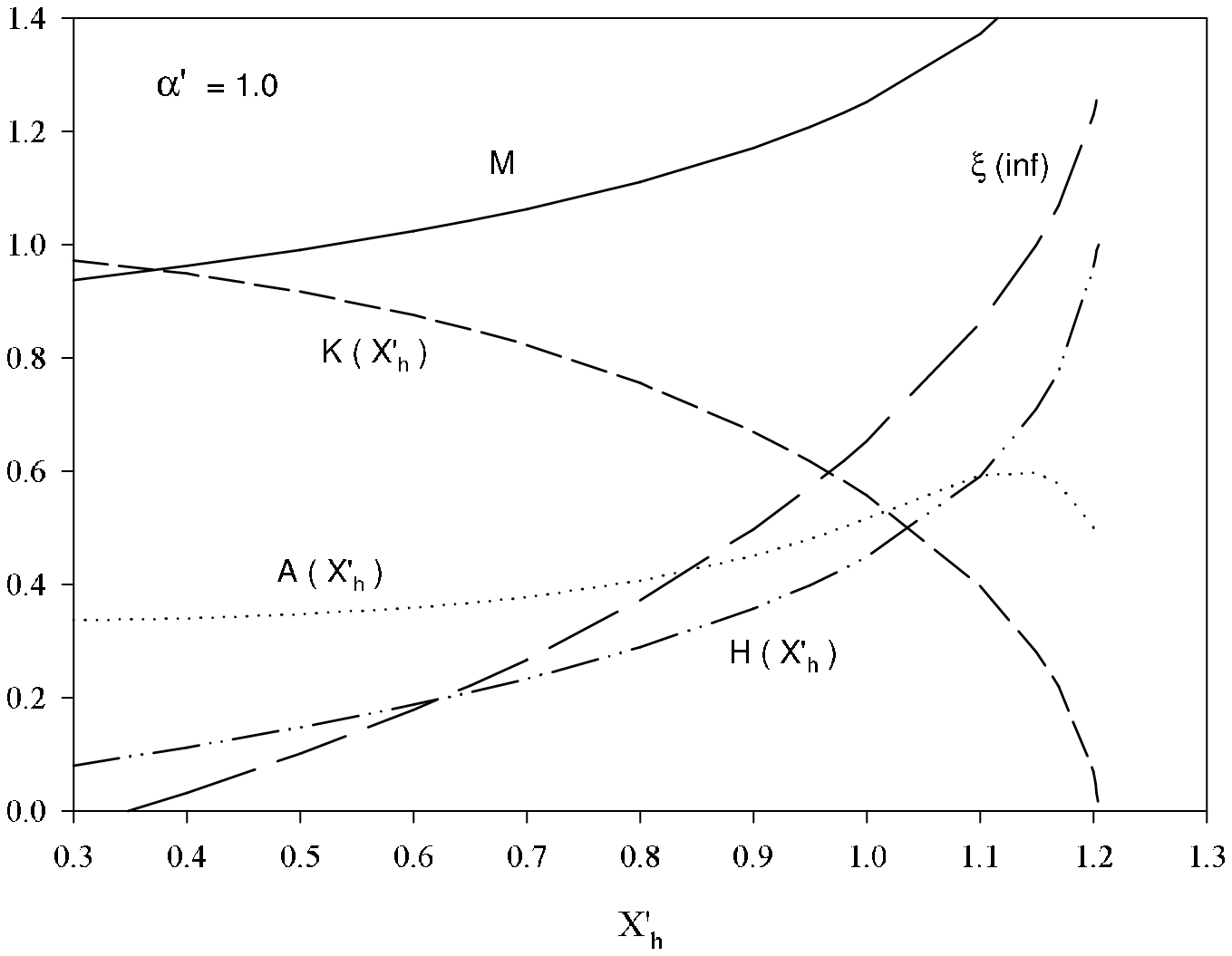,width=12cm}}
\end{picture}
\\
\begin{center}
\small{{\bf Fig 5} The parameters characterizing the black string
as functions of the horizon for $\alpha'=1$}
\end{center}
\newpage
\newpage
\setlength{\unitlength}{1cm}
\begin{picture}(6,9)
\centering
\put(1.6,0){\epsfig{file=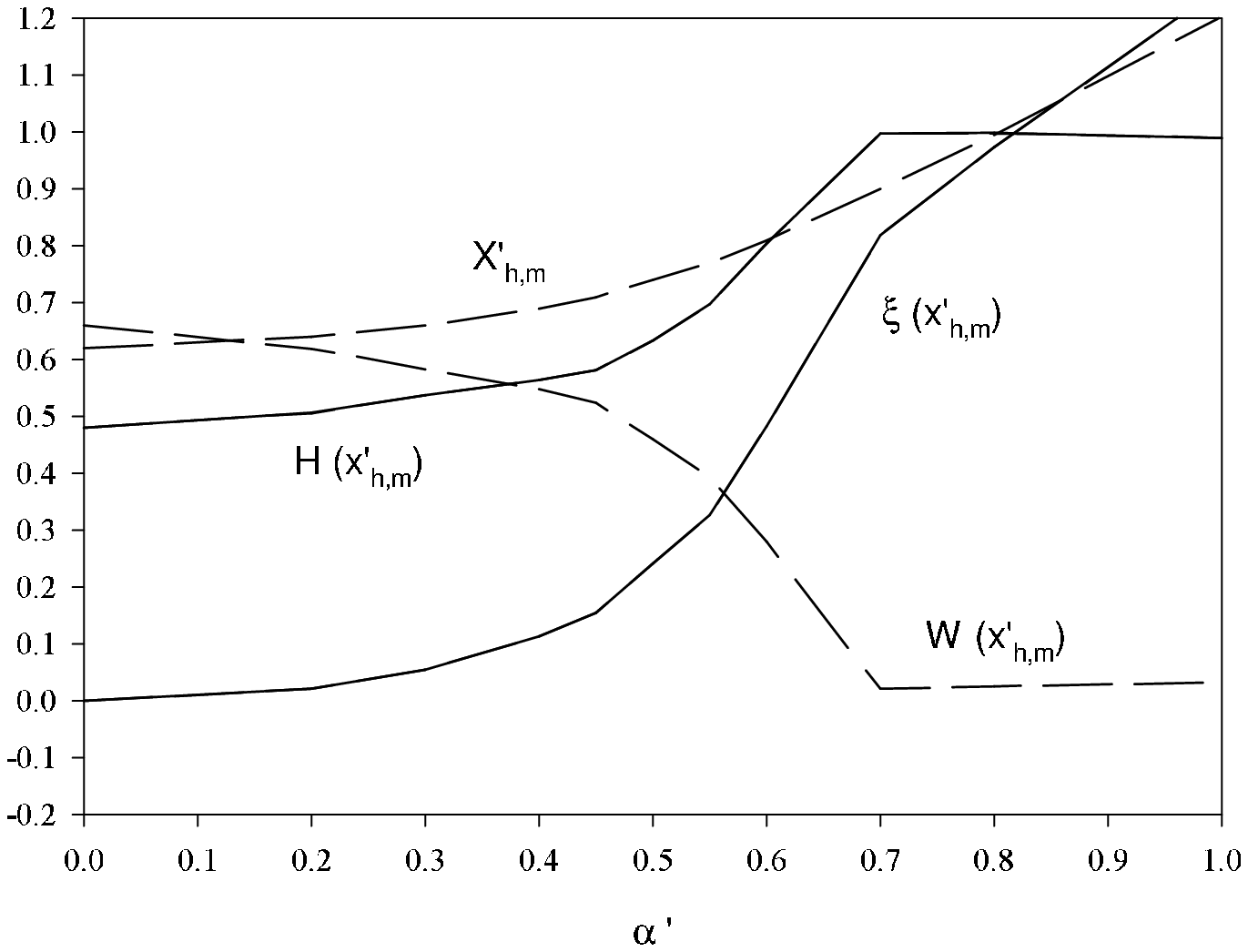,width=12cm}}
\end{picture}
\\
\begin{center}
\small{{\bf Fig 6} The maximal value of the horizon $x'_{h,max}$
is plotted as a function of $\alpha'$
a few parameters related to the transition are supplemented}
\end{center}
\newpage
\begin{center}
\end{center}
\begin{picture}(10,7.7)
\centering
\put(2.2,0){\epsfig{file=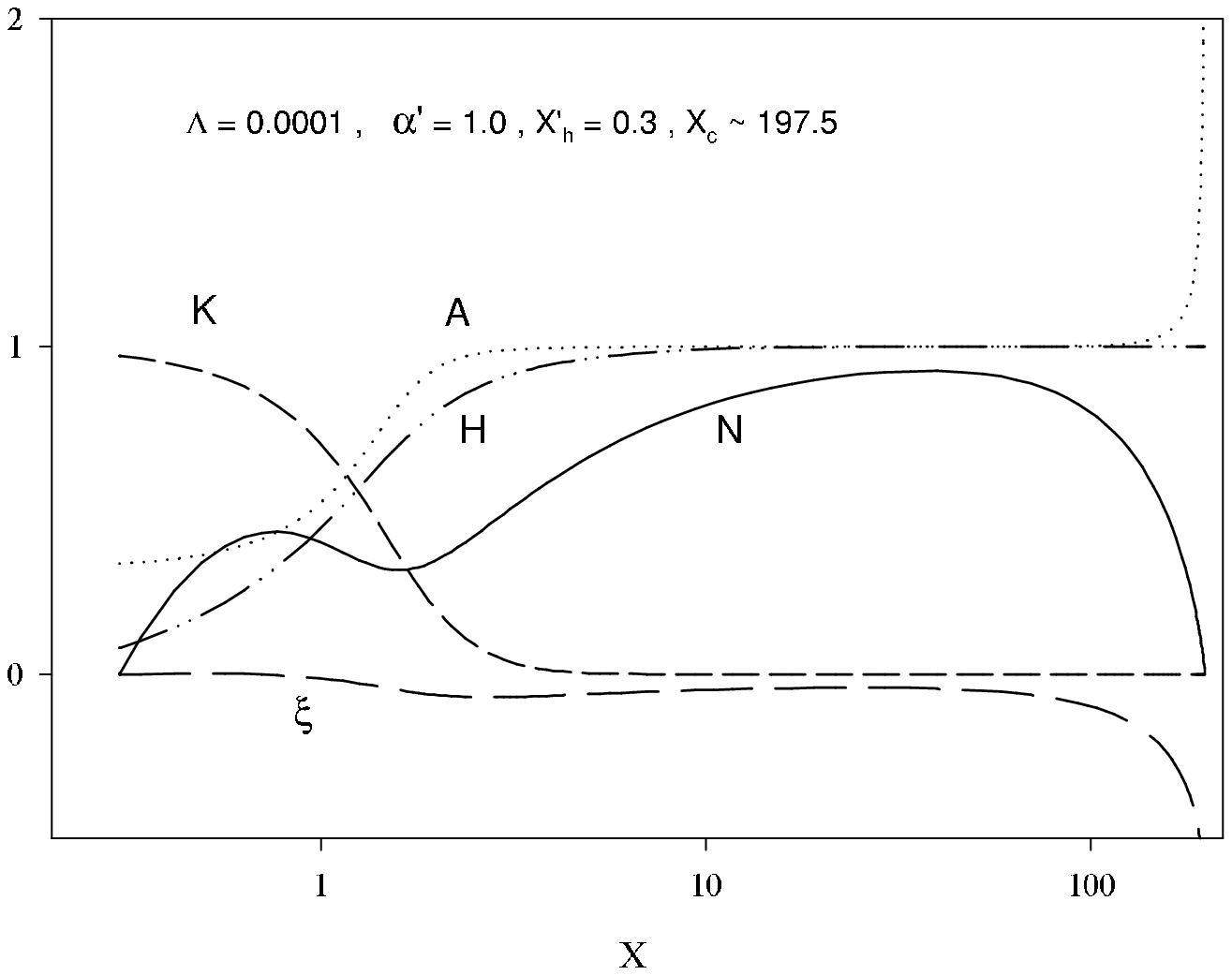,width=12cm}}
\end{picture}
\begin{center}
\small{{\bf Fig 7} 
The profiles of a black string with cosmological
constant $\Lambda = 0.0001$}
\end{center}

\newpage
\begin{figure}
\centering
\epsfysize=18cm
\mbox{\epsffile{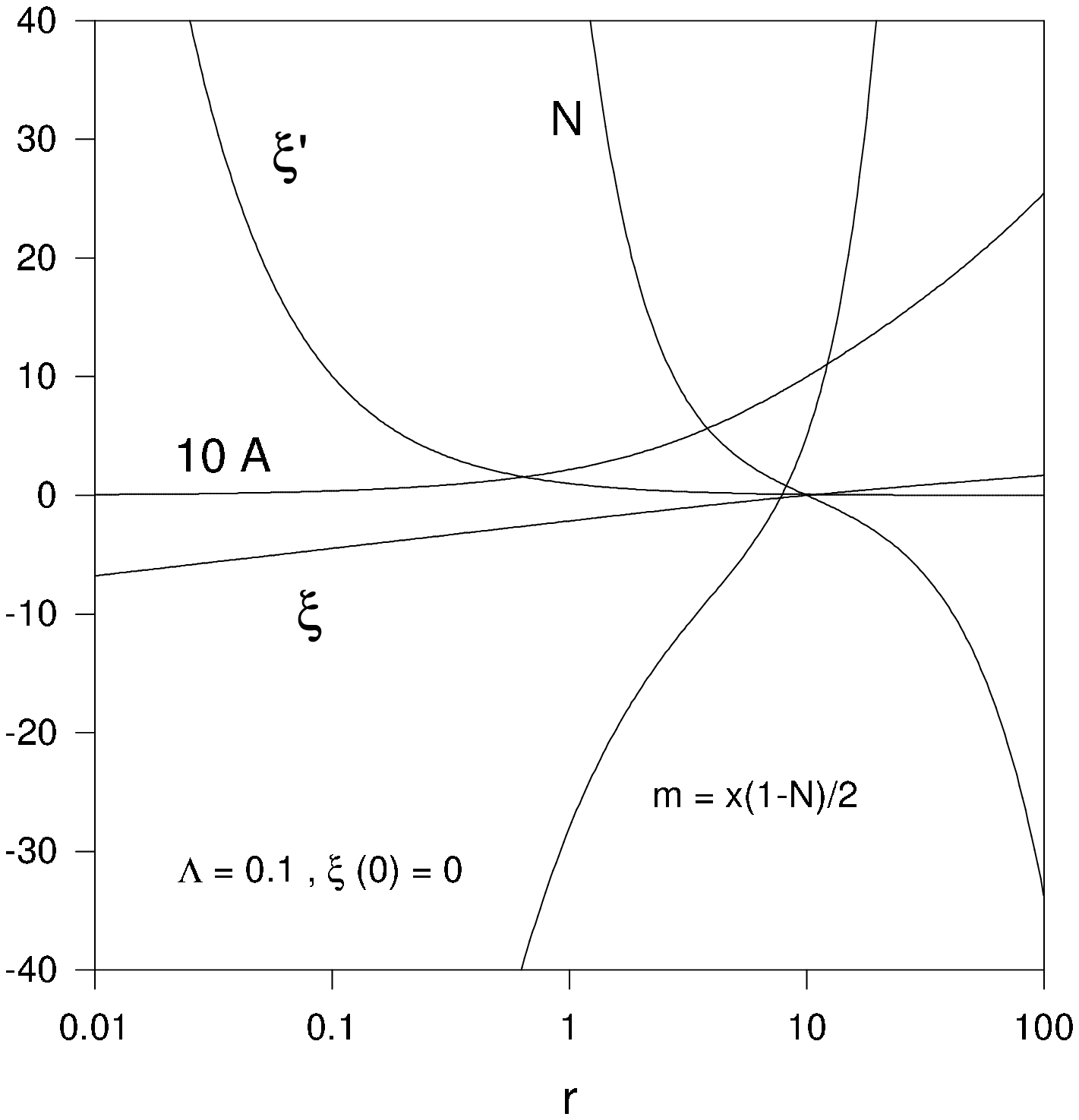}}
\begin{center}
\small{{\bf Fig 8}
The values of the mass, dilaton charge, $A(0), N_{min}$
and $\xi(\infty)$ are plotted as functions of $\alpha_c$
for the solutions with $\xi(0)=0$}
\end{center}
\end{figure}


\begin{thebibliography}{99}
\bibitem{vilenkin}A. Vilenkin and E.P.S. Shellard,
{\it Cosmic Strings and Other Topological Defects}
Cambridge University press (1994).
\bibitem{thooft}
 G. `t Hooft, 
 Nucl.~Phys.~ {\bf B 79} (1974), 276;\\
 A.~M. Polyakov, 
 JETP Lett. {\bf 20} (1974), 194.

\bibitem{astrocc}
  S.~Perlmutter {\it et al.}  [Supernova Cosmology Project Collaboration],
  Nature {\bf 391} (1998) 51
  [arXiv:astro-ph/9712212];
\newline
  A.~G.~Riess {\it et al.}  [Supernova Search Team Collaboration],
  Astron.\ J.\  {\bf 116} (1998) 1009
  [arXiv:astro-ph/9805201].
\bibitem{bhrds}
Y. Brihaye, B. Hartmann and E. Radu, Phys. Rev. Lett. {\bf 96} (2006),0701101.
\bibitem{brane} K. Akama, ``Pregeometry'',
in Lecture Notes in Physics, 176, Gauge Theory
and Gravitation, Proceedings, Nara, 1982, edited by K. Kikkawa, N. Nakanishi
and H. Nariai, 267-271 (Springer-Verlag, 1983) [hep-th/0001113];
V. A. Rubakov and M. E. Shaposhnikov, Phys. Lett. {\bf 125 B} (1983), 136; {\bf 125 B} (1983), 139;
G. Davli and M. Shifman, Phys. Lett. {\bf B 396} (1997), 64; {\bf 407} (1997), 452;
I. Antoniadis, Phys. Lett. {\bf B 246} (1990), 377; 
N. Arkani-Hamed, S. Dimopoulos and G. Dvali, Phys. Lett. {\bf B 429} (1998), 263;
I. Antoniadis, N. Arkani-Hamed, S. Dimopoulos and G. Dvali, Phys. Lett. 
{\bf B 436} (1998), 257; L. Randall and R. Sundrum, Phys. Rev. Lett. {\bf 83} (1999), 3370; {\bf 83} (1999), 4690. 
\bibitem{tan}  F. R. Tangherlini, Nuovo Cimento {\bf 27} (1963), 636.
\bibitem{mp} R. C. Myers and M. J. Perry, Ann. Phys. (NY) {\bf 172} (1986), 304.
\bibitem{gl} R. Gregory and R. Laflamme, Phys. Rev. {\bf D 37} (1988), 305.
\bibitem{er} R. Emparan and H. Reall, Phys. Rev. Lett. {\bf 88} (2002), 101101.
\bibitem{bcht}
Y. Brihaye, A. Chakrabarti, B. Hartmann and D. H. Tchrakian,
Phys. Lett. {\bf B 561} (2003), 161.

\bibitem{volkov}
M. S. Volkov, Phys. Lett. {\bf B 524} (2002), 369.
\bibitem{bh1} Y. Brihaye and B. Hartmann, Phys. Lett. {\bf B 534} (2002), 137.
\bibitem{hartmann}
  B.~Hartmann,
  Phys.\ Lett.\ B {\bf 602} (2004) 231
  [arXiv:hep-th/0409006].
\bibitem{bhr}
  Y.~Brihaye, B.~Hartmann and E.~Radu,
  arXiv:hep-th/0508028, Phys. Rev. {\bf D 72} (2005) 104008;
  arXiv:hep-th/0502131, Phys. Rev.  {\bf D 71} (2005) 085002.
\bibitem{bh2}
  Y.~Brihaye and B.~Hartmann,
  arXiv:gr-qc/0503102, Class. Quant. Grav. {\bf 22} (2005) 5145.

\bibitem{bch} Y. Brihaye, F. Clement and B. Hartmann,
arXiv:hep-th/040603],   Phys. Rev. {\bf D70 }, (2004) 084003.
\bibitem{bfm} P. Breitenlohner, P. Forgacs and D. Maison,
Nucl. Phys. {\bf B 383} (1992),357; Nucl. Phys. {\bf B 442} (1999),
126.
\bibitem{bbh} B. Hartmann, Y. Brihaye and B. Bertrand,
Phys. Lett. {\bf B 570} (2003), 137.
\bibitem{mann} K.C.K. Chan, J.H. Horne and R.B. Mann,
Nucl. Phys. B {\bf 447} (1995) 441.
\bibitem{mrs} R.B. Mann, E. Radu and C. Stelea, "Black string solutions
with negative cosmological constant", arXiv:hep-th/0604205.
\bibitem{brs} Y. Brihaye, E. Radu and C. Stelea, in preparation.



\end{thebibliography}
\end{document}